\documentclass{aa}  
\usepackage{graphicx}
\usepackage{txfonts}
\usepackage{natbib}
\usepackage{caption}
\usepackage{hyperref}
\bibpunct{(}{)}{;}{a}{}{,}

\begin{document} 

	\title{Astrokit---an Efficient Program for High-Precision Differential CCD Photometry and Search for Variable Stars}

	\author{Artem Y. Burdanov\thanks{e-mail: burdanov.art@gmail.com}, Vadim V. Krushinsky, \and Alexander A. Popov}

	\institute{Ural Federal University, Ekaterinburg, 620002, Russia}

	\date{Received December 24, 2013; accepted May 28, 2014}

 
	\abstract
	{Having a need to perform differential photometry for tens of thousands stars in a several square degrees field, we developed {\tt Astrokit} program. The            	software corrects the star brightness variations caused by variations of atmospheric transparency: to this end, the program selects for each star an  				individual ensemble of reference stars having similar magnitudes and positions in the frame. With ten or more reference stars in the ensemble, the 				differences between their spectral types and the spectral type of the object studied become unimportant. {\tt Astrokit} searches for variable stars using 			Robust Median Statistics criterion, which allows candidate variables to be selected more efficiently than by analyzing the standard deviation of star 				magnitudes. The software allows very precise automatic analysis of long inhomogeneous sets of photometric observations of a large number of objects to be 			performed, making it possible to find ``hot Jupiter''\ type exoplanet transits and low-amplitude variables. We describe the algorithm of the program and the 		results of its application to reduce the data of the photometric sky survey in Cygnus as well as observations of the open cluster NGC\,188 and the transit of 	the exoplanet WASP-11\,b\,/\,HAT-P-10\,b, performed with the MASTER-II-URAL telescope of the Kourovka Astronomical Observatory of the Ural Federal 				University.}


	\titlerunning{Astrokit}
	\authorrunning{Burdanov et al.}
	\maketitle

	\section{Introduction}

	We developed {\tt Astrokit} C++ console application for post-processing of the results of CCD photometry within the framework of a program for the search for 	new variable stars and exoplanet transits in the Kourovka Astronomical Observatory of the Ural Federal University. The application is based on an upgraded
	algorithm of differential photometry using ensembles of comparison stars as described in~\citet{Everett2001}.

	In this paper we do not consider the sources of errors that influence the precision of CCD photometry. For a detailed discussion and analysis of this problem 	and the use of other methods of CCD photometry see~\citet{Howell1988, Gilliland1988, Gilliland1992, Newberry1991, Young1991, Honeycutt1992,  						Gilliland1993, Merline1995, Howell2000, Everett2000}. We only recall that the idea of differential photometry consists in determining the difference between 		the magnitude of the source studied and that or those of 	one or several reference stars, thereby reducing the influence of time-variable atmospheric effects. 	In the ideal case the reference star should have similar 	brightness, color index, and must be located close to the star studied. 

	The close location of the reference stars to the star studied is 	especially important in the case of the reduction of wide-field images. This is because 			otherwise local variations of atmospheric transparency would have different effect on the reference sources and on the object studied, thereby inevitably 			degrading the resulting photometric precision.

	After performing differential photometry {\tt Astrokit} searches for variable stars. Below we describe the algorithm of the program, its implementation, and 		some of the results 	obtained.

	\section{Method}

	Before starting post-processing of the photometric data, photometry proper has to be performed. It makes no difference what software was used to extract 			fluxes from the CCD frame, the only 	important thing is to obtain the corresponding magnitudes and their theoretical errors in accordance with the main CCD
	equation~(\citet{Howell1993}).

\begin{figure}[t]
	\centering	
	\includegraphics[width=0.4\columnwidth]{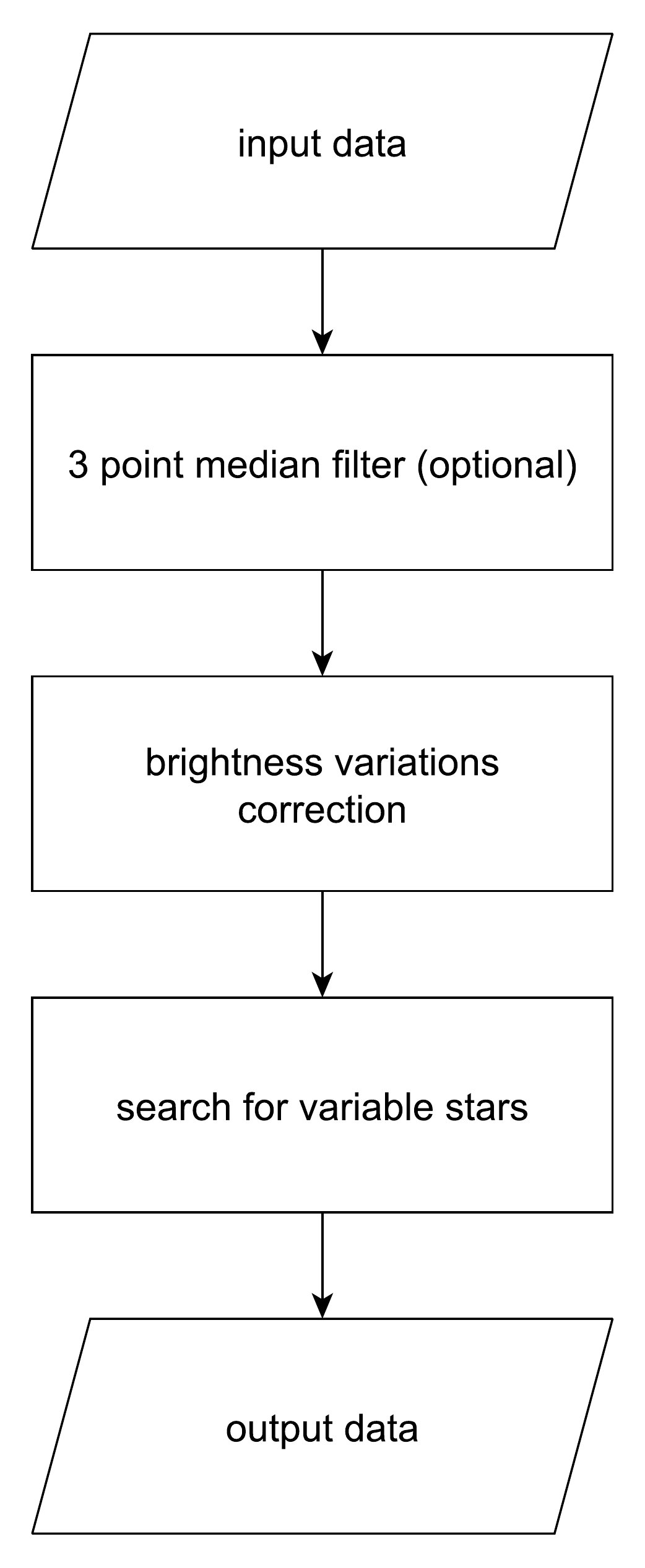}
 	\captionsetup{justification=centering}
	\caption{Structure of {\tt Astrokit} program.}
	\label{fig01}
	\end{figure}

	For the studies carried out at the Astronomical Observatory of Ural Federal University, a dedicated technique of photometric reduction was developed within 		the framework of IRAF package~(\citet{Tody1986}).\linebreak Before applying IRAF the console version of\linebreak {\tt Astrometry.net} 							application~(\citet{Lang2010}) is used to set the correct World Coordinate System parameters in the FITS header of each frame. IRAF package is 	then 			used to perform photometric reduction of each frame:  subtraction of the dark frame and dividing by the flat-field frame. The  {\tt PHOT/APPHOT} task is 			then used to perform aperture photometry in each frame with individual aperture and sky background values for each frame. A catalog of objects
	containing the equatorial coordinates and running numbers of stars is used to this end. The aperture radius used in each particular frame is set equal to 			$0.8\,{\rm FWHM}$, where $\rm FWHM$ is the mean  full width at half maximum value of the stellar PSF in the frame. The resulting data are then transferred to  	{\tt Astrokit}, 	whose general structure is shown in Fig.~\ref{fig01}.
	
	Input data for the program are contained in the file created by {\tt PDUMP} command of IRAF package and include:
	\begin{list}{--}{
	\setlength\leftmargin{5mm} \setlength\topsep{2mm}
	\setlength\parsep{0mm} \setlength\itemsep{2mm} }
  	\item identification number of the star ({\tt id}),
  	\item number of counts (analog-to-digital units) from the star inside the aperture together with the sky background counts ({\tt sum}),
  	\item aperture area in square pixels ({\tt area}),
  	\item average sky background in each pixel ({\tt msky}),
  	\item number of pixels attributed to the sky background ({\tt nsky}),
  	\item exposure ({\tt itime}).
	\end{list}

	For the program to operate correctly, it also needs a file containing the equatorial coordinates of the stars. This file may also optionally contain the 			catalogued color indices which will be later taken into account when selecting ensembles of comparison stars.

	Below we give a stage-by-stage description of the algorithm. Let the available set of photometric data consist of $j$ CCD frames each  containing  $i$ stars 		from the input catalog.

	\vspace{5pt} \noindent (1) The program computes the magnitudes ({\tt m}) and magnitude errors ({\tt merr}) for each star of the input catalog and each frame. 	These quantities are determined as follows:
	\begin{equation}
	\begin{array}{lcl}
 	{\tt flux} &=& {\tt sum} - {\tt area} \times {\tt msky}, \\[+10pt]
 	{\tt m} &=& {\tt zmag} - 2.5\log({\tt flux}) + 2.5\log({\tt itime}), \\[+10pt]
 	{\tt merr} &=& \displaystyle\frac{1.0857}{{\tt flux} \times {\tt gain}}\, \Bigl({\tt flux} \times {\tt gain}\\
 	\lefteqn{+\,{\tt area}\,\bigl(1+{\tt area}/{\tt nsky}\bigr)\,\left({\tt msky} \times
 	{\tt gain} + {\tt ron}^2\right)\Bigr)^{\frac{1}{2}}\!\!\!,} & & \\[+5pt]
	\end{array}
	\end{equation}
	where ${\tt zmag} = 20$ is the zero point of the magnitude scale; {\tt gain} is the CCD gain in \mbox{e$^-$/ADU}, which is set by 	the user when beginning to 		work with the program; {\tt ron} is the CCD readout noise in e$^-$/pixel, which is also set by the user.
	
	\vspace{5pt} \noindent (2) For each star from the input catalog, an ensemble of reference stars located within a certain radius (the default value 				is~$5\arcmin$) is selected whose magnitudes differ from that of the star considered by no more than $2^{\rm m}$ (this parameter  can be changed by the user). 	The smaller the magnitude difference, the smaller the number of stars in the ensemble, and the farther they are. Another criterion for selecting ensemble 			stars is based on the difference between the color index of the star considered  (i.e., the star for which the ensemble is composed) and candidate comparison 	stars.
	
	\vspace{5pt} \noindent (3) The weighted average instrumental magnitude $\langle m_j\rangle$ of ensemble stars is computed for each frame of the series
	$$\langle m_j\rangle = \frac{\sum\limits_k^K m_{kj} \, \omega_k}{\sum\limits_k^K \omega_k},\quad\mbox{where}\quad \omega_k=\frac{1}{\langle {\tt 					merr}_k^2\rangle},$$ where $k$ is the running number of the star in the ensemble, $K$ is the number of stars in the ensemble, $j$ is the frame number,
	$m_{kj}$ is the magnitude of \mbox{$k$-th} ensemble star in $j$-th frame, $\langle {\tt merr}_k\rangle$ is the mean theoretical error of measured magnitude 		$m_k$ of the ensemble star computed by formula~(1) (instead of the error in each frame as in~\citet{Everett2001}).

	\vspace{5pt} \noindent (4) The mean magnitude $M$ of all ensemble stars averaged over all frames: $$M = \frac{\sum\limits_j^N \langle m_j\rangle}{N},$$
	where $N$ is the number of frames.

	\vspace{5pt} \noindent (5) The difference between the weighted average magnitude $\langle m_j\rangle$ of ensemble stars and the mean magnitude $M$ of all 			ensemble stars averaged over all frames is then subtracted from the observed magnitude of the star for which the effect of the terrestrial atmosphere is 			determined and for the ensemble stars: $$m_{{\rm cor}\,ij} = m_{ij} - (\langle m_j\rangle - M),$$ where $m_{{\rm cor}\,ij}$ and $m_{ij}$ are respectively the
	corrected and initial magnitudes of star $i$ in frame $j$.

	\vspace{5pt} \noindent (6) After composing the initial ensemble, the standard deviation from the mean magnitude is computed for all stars, and the star with 		the greatest standard deviation is flagged. If the standard deviation from the mean magnitude is more than twice greater than the mean theoretical  				photometric error averaged over all frames (we call this the cutoff ratio of the sigma criterion), which can also be varied, the star is removed from the 			ensemble, and the procedure is repeated from step~2.

	If after all stars with large standard deviations are removed the ensemble contains less than 10 stars, the size of the ensemble domain is increased by  			$1\arcmin$ and all the above steps are repeated. The correction of instrumental magnitudes is thus an iterative process, which is repeated
	until the ensemble contains more than nine stars or the search radius increases to  $30\arcmin$.

	The error introduced by the correction of the initial magnitudes using the ensemble stars is determined by the magnitude errors of  the ensemble stars:
	$${\tt merr}_{{\rm ens}} = \frac{1}{\sqrt{\sum\limits_{k=1}^K \displaystyle\frac{1}{{\tt merr}_k^2}}},$$ where ${\tt merr}_k$ is the measurement error for 		$k$-th star of the ensemble. The resulting error for star $i$ is composed of its measurement error and the error introduced by the comparison
	ensemble: $${\tt merr} = \sqrt{{\tt merr}_{i}^2 + {\tt merr}_{\rm ens}^2}.$$

	The procedure of the formation of ensembles and the correction of instrumental magnitudes is performed for each star of the list. Thus, for each star its 			individual ensemble of closely located comparison stars is created.

\begin{figure}[h]
	\centering
	\includegraphics[width=0.7\columnwidth]{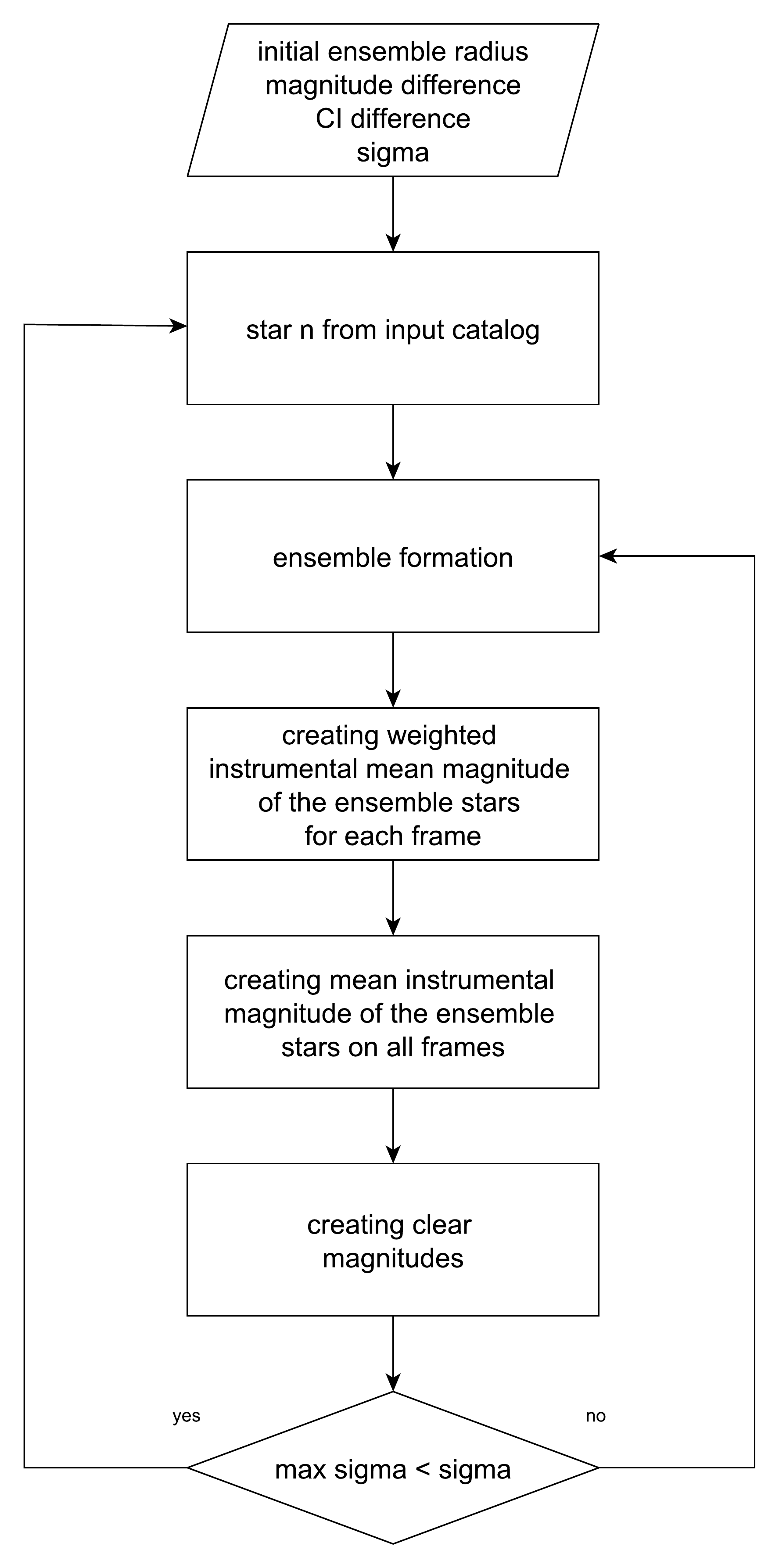}
	\caption{Flowchart of the process of the correction of instrumental magnitudes.} \label{fig02}
	\end{figure}

	Figure~\ref{fig02} shows schematically the process of the correction of instrumental magnitudes.

	\vspace{5pt} \noindent (7) Variable objects are identified by the algorithm described in~\citet{Rose2007}. For each star 
	the coefficient  ${\rm RoMS}$ (Robust Median Statistics) is computed: $$\eta_n = \frac{\sum\limits_i\displaystyle\frac{\mid m_i - \langle m_{\rm 					med}\rangle\mid}{\sigma_{\rm rms}}}{N-1},$$ where $n$ is the number of a star; $m_i$ is the $i$-th measurement of the magnitude; $\langle m_{\rm med}\rangle$ 	is the median of magnitude measurements of star $n$; $N$ is the total number of magnitude measurements for star $n$; $\sigma_{\rm rms}$ is the
	estimated standard deviation of star $n$, determined by the least-squares method from the dependence of the standard deviations of the magnitude on the 			stellar magnitude for stars in the frame.

	The ${\rm RoMS}$ criterion allows estimating the variations of the object brightness. If it exceeds $1$, the star is considered to be a suspected variable 		and is then analyzed more thoroughly.
	\vspace{5pt}

	\section{Analysis of the technique}
	
	To analyze the technique and select the optimum input parameters for {\tt Astrokit} to ensure the best photometric precision, we used the data of the 				photometric survey of a sky area in the Milky Way. A total of 3000 50-second $R$-band frames were taken with the MASTER-II-URAL telescope. In the central 			\mbox{$30'\times30'$} region of the frames, we selected 800~stars in the $R = 9^{\rm m}$--$17^{\rm m}$ magnitude interval.

	The MASTER-II-URAL is located at the Kourovka Observatory of the Ural Federal University. It is a Hamilton system twin telescope ($D = 40$~cm, $F = 				1000$~cm) installed on equatorial mounting and equipped with two Apogee Alta~U16M CCD cameras~(\citet{Lipunov2010}). The image scale is
	$1\farcs8$/pixel. Photometric calibrations are performed using dark frames taken before each observing night and dawn flat-field frames. All observations are 	performed in automatic mode.

	After performing aperture photometry with IRAF, we carried out a series of reduction cycles with {\tt Astrokit}. We varied such input parameters as the 			initial radius $r$ of the ensemble, the magnitude difference $\Delta m$ and color-index difference $\Delta{\rm CI}$ of ensemble stars, and the cutoff ratio
	($\sigma$) of the standard deviation of instrumental magnitude to the theoretical error (the sigma criterion).

	We consider the main criterion characterizing the quality of post-processing to be the number of ``good stars,\!'' i.e., stars with the standard deviation of 	magnitudes $s$ of less than $0\fmm01$ and $0\fmm02$ over the entire observing set. We also took into account the minimum computed standard deviation of 			magnitude for an individual star (hereafter the ``best star'').

	We first varied the initial radius of the ensemble with the magnitude difference fixed at $1^{\rm m}$ (shown by the squares in plots) and the  sigma 				criterion equal to two. The initial ensemble radii were equal to  $1'$, $2'$, $3'$, $4'$, $5'$, $7'$, $10'$, and $15'$. We then counted for each case the 			number of ``good stars''\ and the minimum standard deviation for each initial radius.

\begin{figure}[t]
	\centering
	\includegraphics[width=0.9\columnwidth]{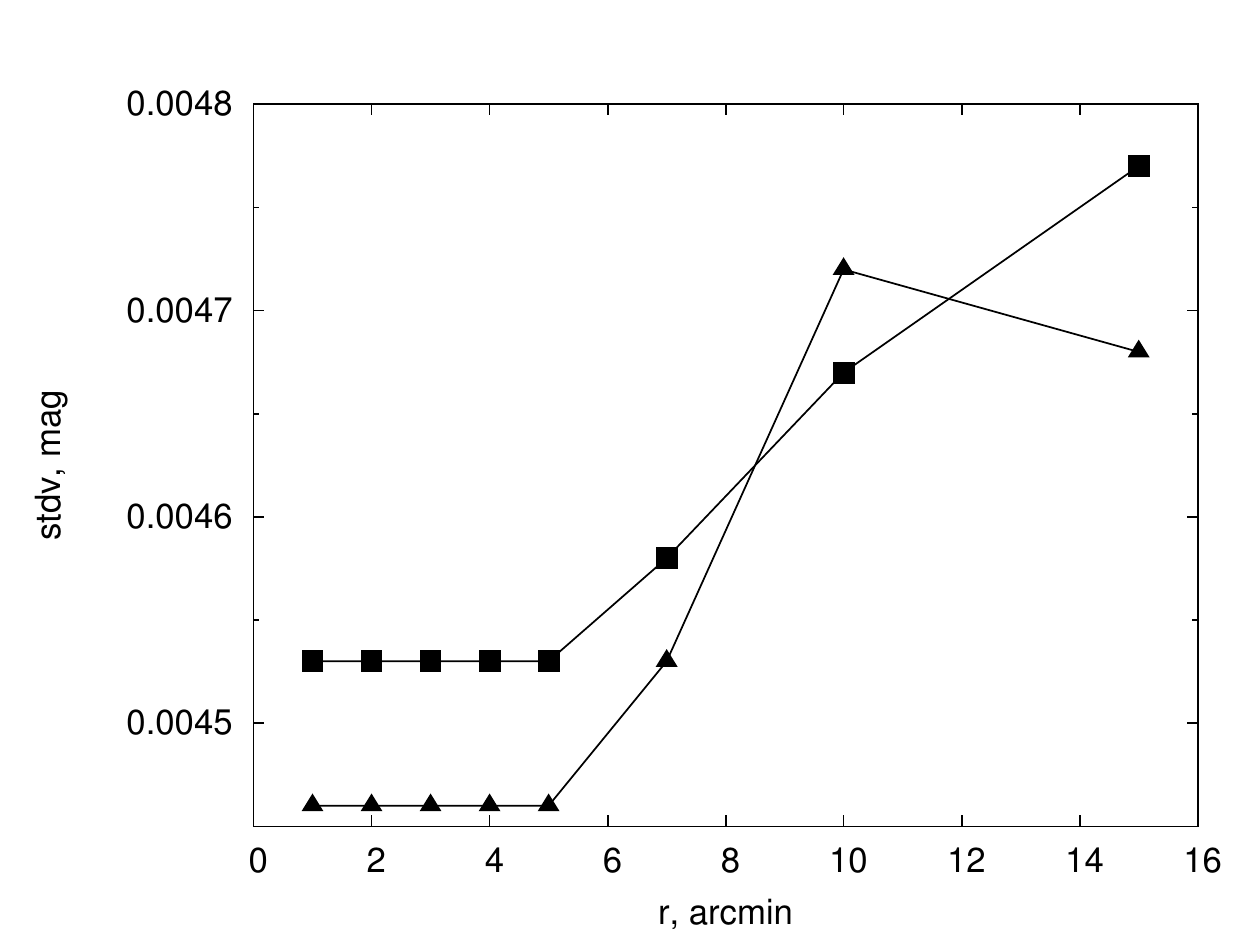}
	\caption{Variation of the standard deviation from the mean magnitude for the ``best star''\ as a function of the initial ensemble radius for $\sigma=2$ and 		different  $\Delta m$ values (the dependences for \mbox{$\Delta m=1$} and $\Delta m=2$ are shown by the squares and triangles respectively).}
	\label{fig03}
	\end{figure}

	As is evident from Fig.~\ref{fig03}, the minimum standard deviation from the mean magnitude is equal to $0\fmm00453$ for the initial radii from 		$1'$ 		to $5'$, increases with further increase of the ensemble radius, and reaches the maximum 5\% difference ($0\fmm00477$) for the ensemble radius of 					$15'$.
	
\begin{figure}[t]
	\centering
	\includegraphics[width=0.9\columnwidth, clip]{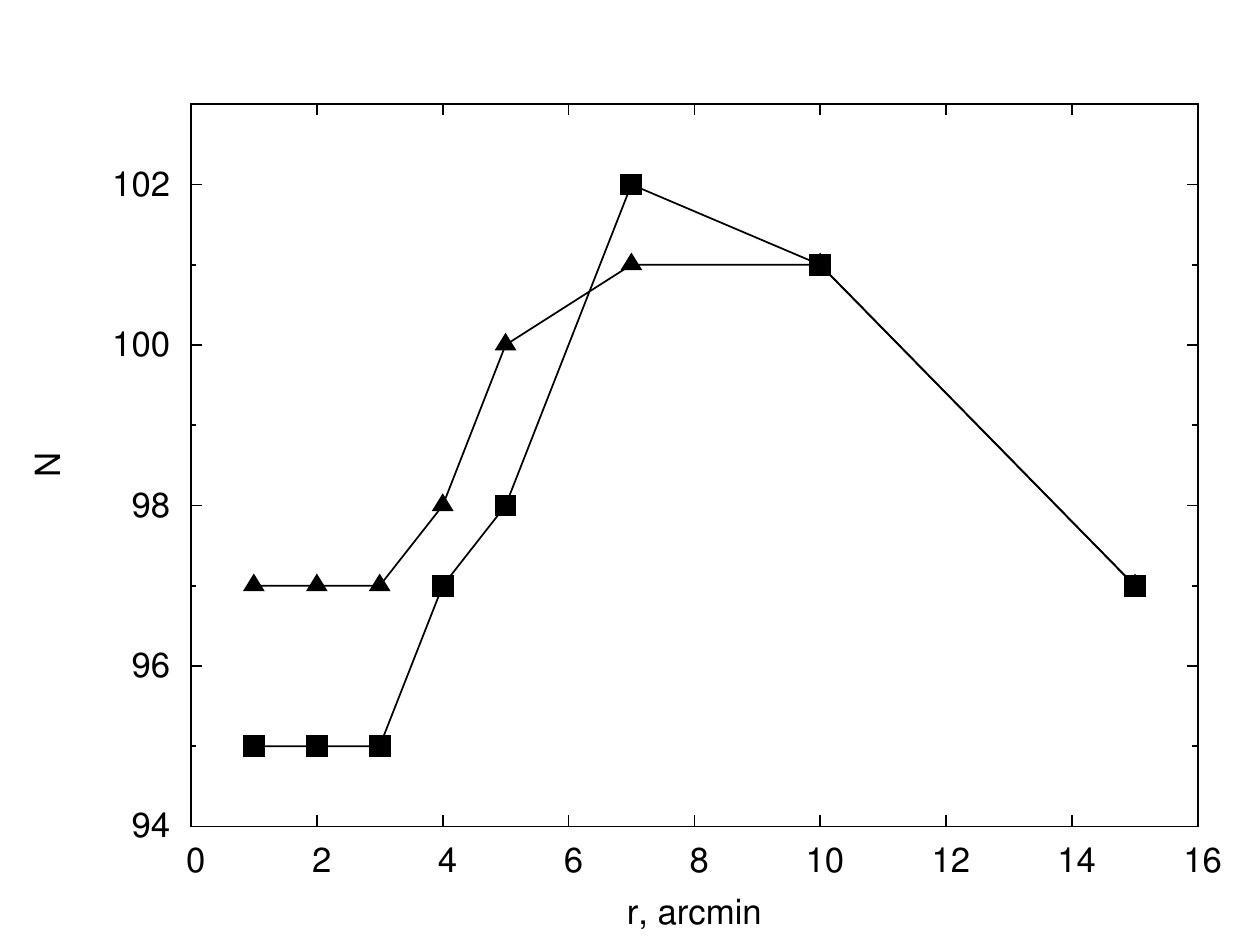}
	\caption{Variation of the number of stars with the standard deviation from the mean magnitude of less than $0\fmm01$ as a 	function of the initial ensemble 		radius for $\sigma=2$ and different $\Delta m$ values (the dependences for \mbox{$\Delta m=1$} and $\Delta m=2$ are shown by the squares and triangles 			respectively).} 
	\label{fig04}
	\end{figure}

\begin{figure}[t]
	\centering
	\includegraphics[width=0.9\columnwidth, clip]{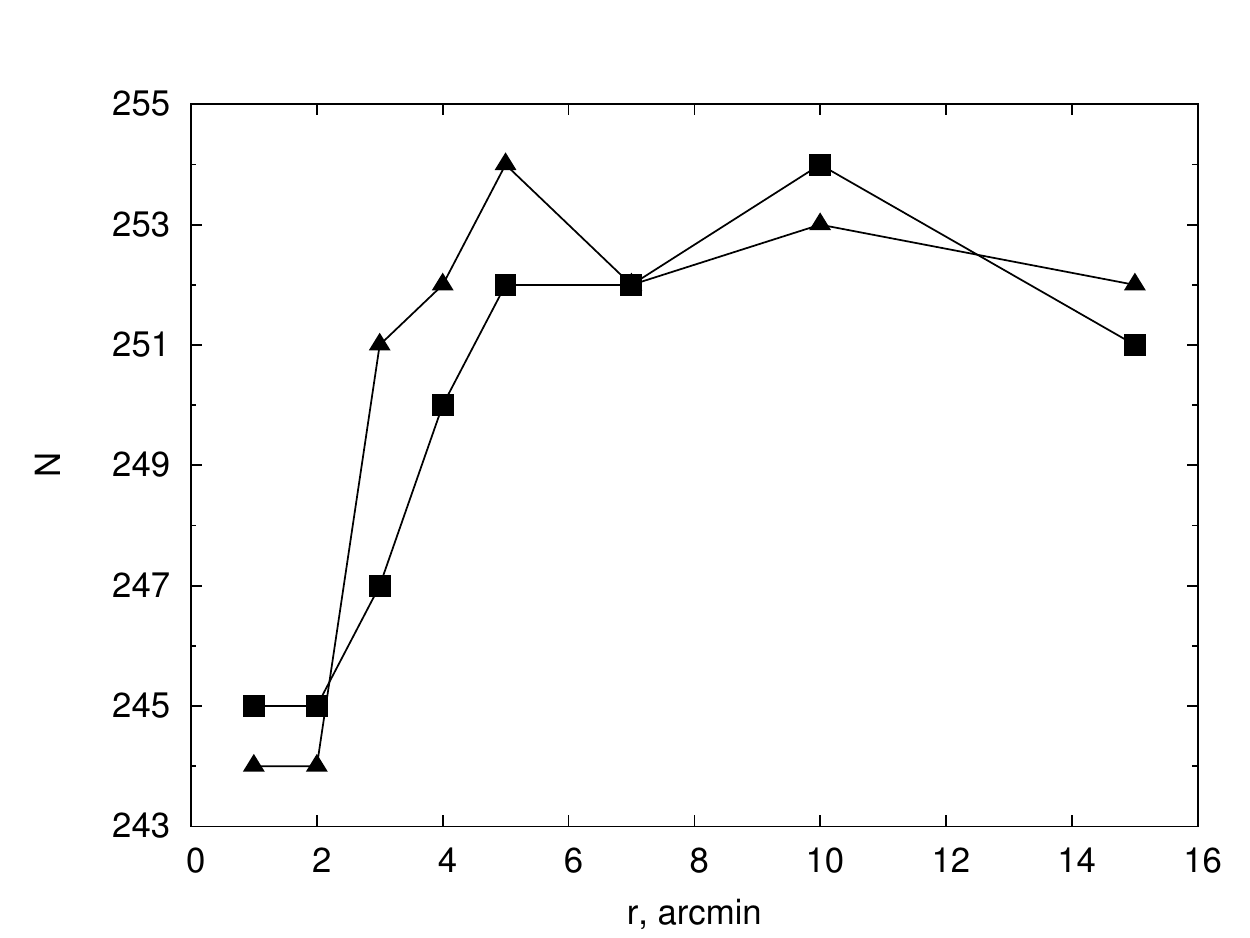}
	\caption{Variation of the number of stars with the standard deviation from the mean magnitude of less than  $0\fmm02$ as a function of the initial
	ensemble radius for $\sigma=2$ and different $\Delta m$ values (the dependences for \mbox{$\Delta m=1$} and $\Delta m=2$ are shown by the squares and 				triangles respectively).}
	\label{fig05}
	\end{figure}

	The number of stars whose standard deviation from the mean magnitude over the entire observing set is less than $0\fmm01$ reaches maximum at the ensemble 			radius of  $7'$: it is equal to 102 which is 7\% greater than the corresponding number of stars for the initial ensemble radii from~$1'$ to~$3'$
	(Fig.~\ref{fig04}). The size of the subsample of stars with the standard deviation from the mean magnitude of less than $0\fmm02$ reaches maximum at the 			initial ensemble radius of  $10'$ (254~stars compared to 245). The increment is equal to 4\% (Fig.~\ref{fig05}).

	\begin{figure}[t]
	\centering
	\includegraphics[width=0.9\columnwidth, clip]{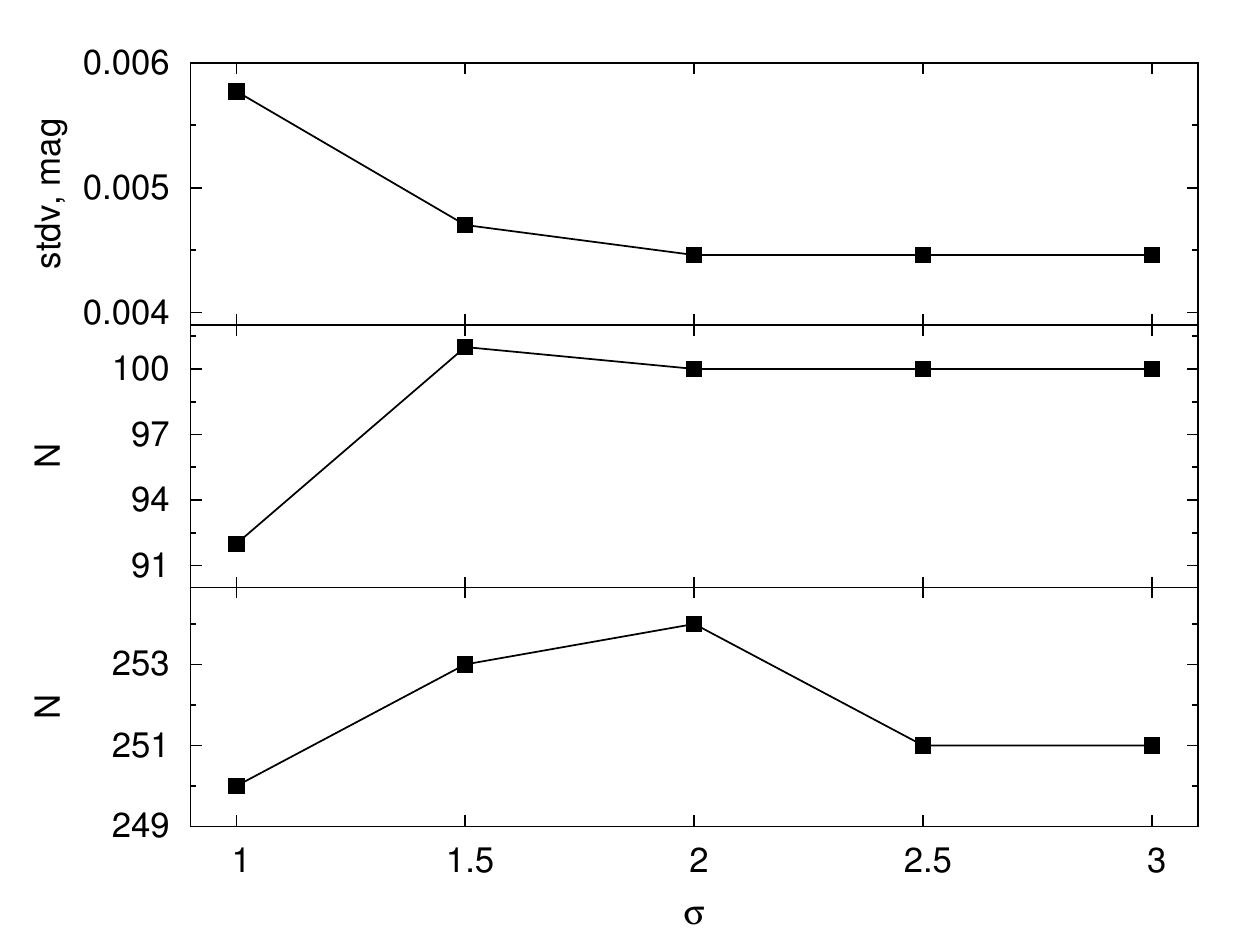}
	\caption{Variation of the standard deviation from the mean magnitude for the ``best star''\ (the top plot) and the number of stars with $s<0\fmm01$ (the 			middle plot), $s<0\fmm02$ (the bottom plot) as a function of the sigma criterion $\sigma$ for $r=5'$ and $\Delta m=2$.}
	\label{fig06}
	\vspace{10pt}
	\end{figure}
	
	We similarly varied the initial ensemble radius with  $\Delta m=2^{\rm m}$ (the corresponding curves in the figures are shown by the triangles). In this case 	the minimum standard deviation from the mean magnitude is  $0\fmm00446$ and also increases with initial radius (Fig.~\ref{fig03}). The number of ``good
	stars''\ with $s<0\fmm01$ reaches maximum at the initial ensemble radii of $7'$~and~$10'$ (it increases by 4\%) (Fig.~\ref{fig04}). The number of  ``good 			stars''\ with \mbox{$s<0\fmm02$} reaches maximum at the initial ensemble radius of $5'$ (it increases by 4\% compared to the minimum value)
	(Fig.~\ref{fig05}).
	
	In view of the above, we can conclude that the optimum initial radius of the ensemble of comparison stars is  $5'$--$7'$ for the magnitude difference of 			$2^{\rm m}$. In this case the ensemble remains sufficiently compact while containing a large number of stars. The ensemble permits atmospheric effects  to be 	reduced, and this reduction can be expressed in terms of the minimum standard deviation from the mean magnitude for the ``best star'' and the maximum number 		of ``good stars.\!''\ Note that the effect of the varied parameters on the final result is relatively small.

	The next stage in the choice of the optimum parameters consisted in varying the sigma criterion with the initial radius and magnitude difference fixed at 			$5'$~and~$2^{\rm m}$ respectively.

	As is evident from Fig.~\ref{fig06}, the optimum value of the sigma criterion is equal to  $2$. This can be explained by the fact that too ``strict'' value 		decreases the number of stars in the ensemble. A sigma criterion value greater than $2$ increases the number of stars in the ensemble by including stars
	with the greatest standard deviation from the mean magnitude over the entire observing set with inevitable effect on the final precision.

	According to postulates of classical differential photometry with a single comparison star and control stars, the best precision can be achieved with a 			comparison star that is most similar to the object studied both in terms of brightness and spectral type. We studied the effect of color index on the 				precision of photometry. To this end, we varied the difference of the 2MASS $J-H$ color indices~(\citet{Skrutskie2006}) when creating the ensemble
	of stars with the initial radius of  $5'$ and the magnitude difference of  $2^{\rm m}$. We set the color-index difference equal to $0\fmm1$, $0\fmm2$, 			$0\fmm3$, $0\fmm4$, $0\fmm5$, $0\fmm6$, and $0\fmm7$.

\begin{figure}[h]
	\centering
	\includegraphics[width=0.9\columnwidth, clip]{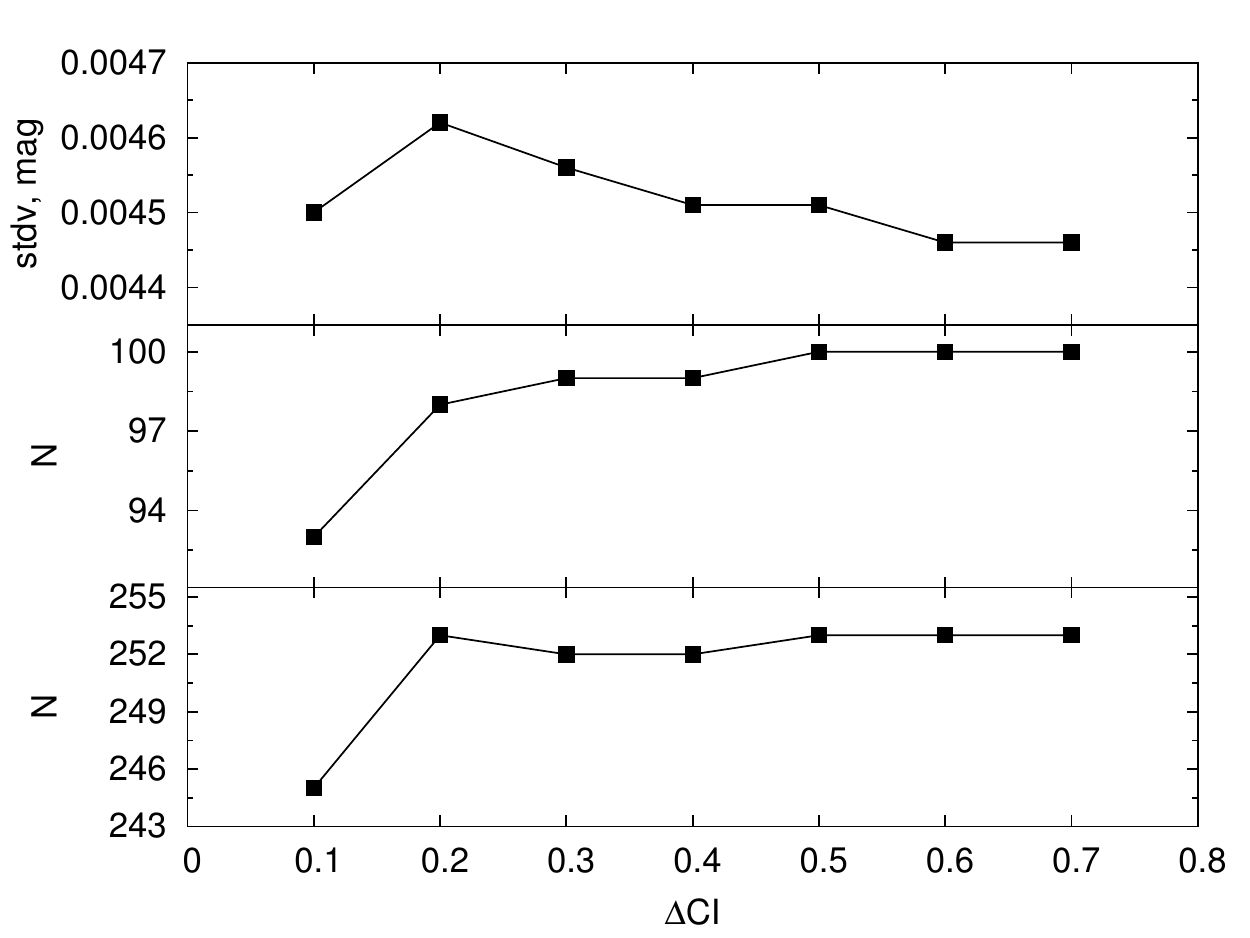}
	\caption{Variation of the standard deviation from the mean magnitude for the  ``best star''\ (the top plot) and the number of stars with $s<0\fmm01$ (the 			middle plot), \mbox{$s<0\fmm02$} (the bottom plot) as a function of the color-index difference $\Delta{\rm CI}$  for $r=5'$ and $\Delta m=2$.}
	\label{fig07}
	\end{figure}

	As is evident from Fig.~\ref{fig07}, the closeness of stellar spectral types is not a necessary condition for achieving high precision. However, the 				classical approach of differential photometry remains  the only solution in the case of small fields and insufficient number of stars for selecting an 			ensemble.

	\begin{figure}[h]
	\centering
	\includegraphics[width=0.9\columnwidth, clip]{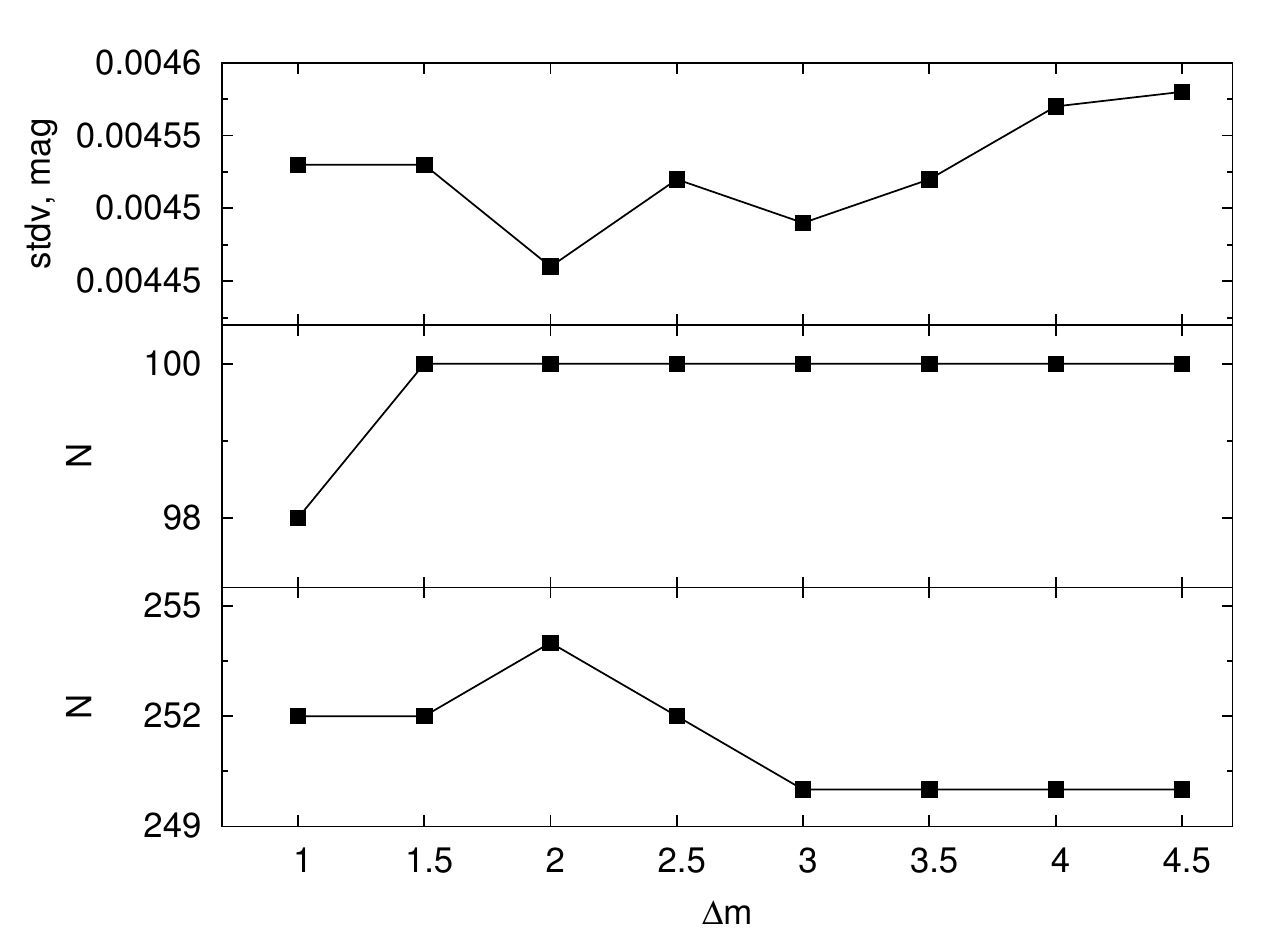}
 	\caption{Variation of the standard deviation from the mean magnitude for the ``best star''\ (the top plot) and the number of stars with $s<0\fmm01$ (the 			middle plot), \mbox{$s<0\fmm02$} (the bottom plot) as a function of the magnitude difference~$\Delta m$ for $r=5'$ and $\sigma=2$.} \label{fig08}
	\end{figure}

	What will happen in the case if the ensemble contains the maximum possible number of reference stars? To this end, we varied the cutoff magnitude for the 			initial ensemble radius and sigma criterion equal to $5'$ and $2$ respectively (Fig.~\ref{fig08}).

	Note that the $2^{\rm m}$ magnitude difference is optimal, because not all possible stars will make it into the ensemble if a smaller value is adopted. 			Adopting a magnitude difference greater than $2^{\rm m}$ would result in the inclusion into the ensemble the stars that are significantly brighter or fainter 	than the object studied, thereby degrading the final precision.

	Our analysis leads us to conclude that in the case of the presence of a sufficient number of stars in the field, the best photometric 	precision is achieved 		using close ensembles with many stars. We thus have the following optimum parameter set for the formation of ensembles of comparison stars: the initial 			ensemble radius $r = 5'$--$7'$, magnitude difference \mbox{$\Delta m=2^{\rm m}$}, and the cutoff ratio of the standard deviation from the mean magnitude
	to the theoretical error for the star to be included into the ensemble  $\sigma=2$. Note that the closeness of the spectral types of stars is of no 				importance.

	At the first sight, small variations in the number of ``good stars''\ and the quality of the  ``best star''\ may appear insignificant and not to be worthy of 	efforts to find the optimum parameters. However, first, there is no harm in increasing the precision. Second, the number of ``good stars''\ may increase
	significantly in the case of reduction of fields containing thousands of stars, and this may influence the number of candidates for searching for transits of 	``hot Jupiter'' type exoplanets  (e.g., if only stars with  $s<0\fmm02$ are selected).

	\section{Testing}

	{\tt Astrokit} program is used for post reduction of photometric data obtained with wide-field telescopes of the MASTER robotic net. Below we report the 			results of its use for reducing a photometric survey of a Milky Way area in Cygnus, and the results of observations of the open cluster  NGC\,188 and the 			transit of the exoplanet WASP-11\,b\,/\,HAT-P-10\,b.

	We reduced all data in accordance with the same procedure. We photometrically calibrated the CCD frames using dark-current and flat-field frames. We observed 	five dark frames for the particular set of exposures at dusk hours before observations. Each master dark frame for the particular exposure is made up of five 	frames via median stacking. We then subtract the master dark frame from the CCD frames with the object studied.

	The flat-field frames in the filter considered are acquired automatically at dawn twilight with the telescope tracking turned off. The linear range of the 		CCDs employed is limited  by  40\,000 ADU, and therefore the exposure in each filter is chosen so that the number of counts in pixels would not exceed this 		limit. The master dark frame with the appropriate exposure is subtracted from each of the five flat-field frames. The flat-field frames are normalized (the 		count of each pixel is divided by the median count of all pixels), and then the master flat-field frame is computed as the median of the normalized initial 		flat-field frames. The CCD frames with the object studied are then divided by the resulting master flat-field frame. The use of flat-field frames allows
	pixel-to-pixel sensitivity variations and vignetting of the optical system to be taken into account.

	After photometric calibration {\tt Astrometry.net} application is used to write the correct astrometric-calibration (WCS) parameters to the FIST file header. 	Aperture photometry is then performed using IRAF {\tt PHOT/APPHOT} task, and the results of this photometry serve as input data for  {\tt Astrokit}.

	\subsection{Photometric Survey in Cygnus}

\begin{figure}[h]
	\centering
	\includegraphics[width=0.9\columnwidth, clip]{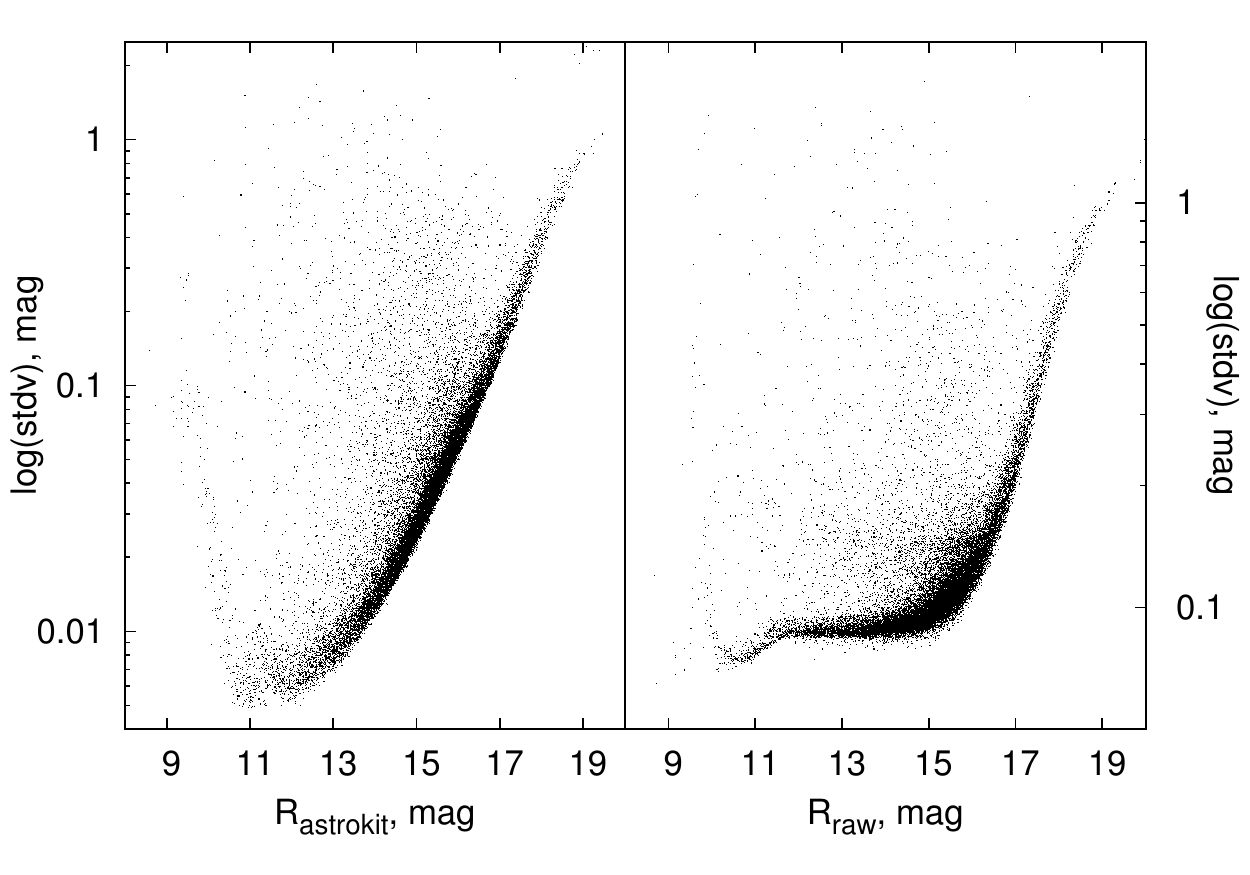}
	\caption{Standard deviation from the mean magnitude as a function of magnitude for the entire observing set: the raw data (right) and the data after 				reduction with {\tt Astrokit} (left).}
	\label{fig09}
	\vspace{-10pt}
	\end{figure}

	High-precision CCD observations were performed at the Kourovka Astronomical Observatory of the Ural Federal University from May through August~2012 within 		the framework of a pilot project aimed at the search for exoplanet transits and variable stars. A total of 3600 frames of of a $2\degr\times2\degr$ area 			centered at \mbox{$\alpha = 20^{\rm h}30^{\rm m}00^{\rm s}$,} \mbox{$\delta = 50\degr30'00''$} were acquired with the MASTER-II-URAL telescope,
	and the photometry of  21\,500~stars was performed with a precision of $0\fmm006$ to $0\fmm5$ in the magnitude interval from $10^{\rm m}$ to $18^{\rm m}$. 		Figure~\ref{fig09} shows the standard deviation from the mean magnitude as a function of magnitude for the entire observing set. The right and left panels
	show the data before and after reduction with {\tt Astrokit}.

	As is evident from the figure, our program decreases the standard deviation from the mean magnitude by a factor of 10 for some stars.

	With the ${\rm RoMS} = 1$ threshold, {\tt Astrokit} selected about 20\%  of all stars as candidate variables. A visual inspection of the light curves of the 		candidate variables allowed us to find 360 hitherto unknown variable stars including:
	\begin{list}{--}{
	\setlength\leftmargin{5mm} \setlength\topsep{2mm}
	\setlength\parsep{0mm} \setlength\itemsep{2mm} }
 	\item 139 stars with periods longer than $20^{\rm d}$;
 	\item 100 stars with periods ranging from  $20^{\rm d}$ to $0\fd1$;
 	\item 19 stars with periods shorter than $0\fd1$;
 	\item 96 eclipsing variables;
 	\item 5 flare stars;
 	\item 1 dwarf nova.
	\end{list}

\begin{figure}[h]
	\centering
	\includegraphics[width=0.9\columnwidth, clip]{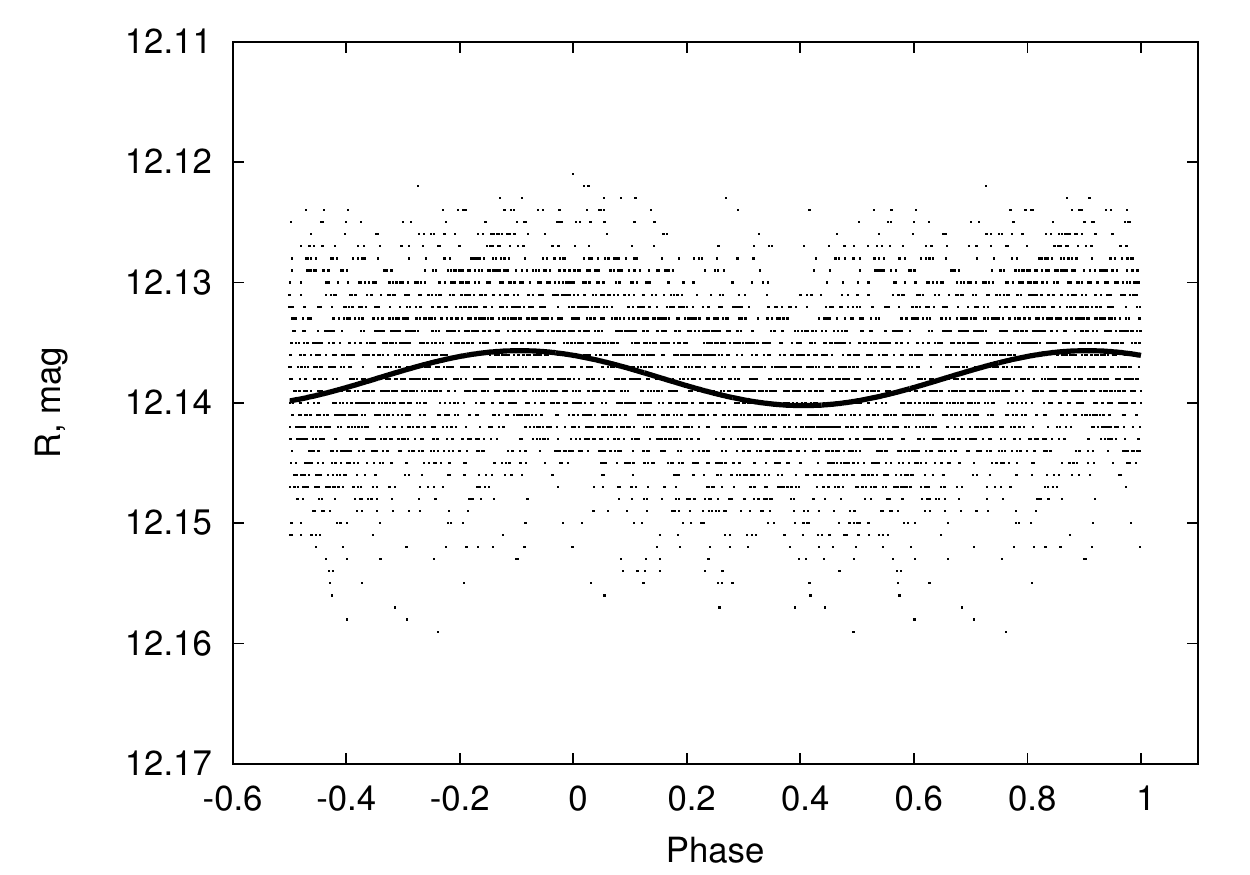}
	\caption{Phased light curve of the star 2MASS 20295743+5017071 with the period and amplitude of $0\fdd035$ and $0\fmm005$ 	respectively. The solid line shows 		the fitted data.}
	\label{fig10}
	\end{figure}

	Among the variables found was the outburst of the dwarf nova USNO-B1.0\,1413-0363790~(\citet{Burdanov2012}). Several dozen variable stars were selected
	as candidate $\delta$\,Sct stars. The brightness of the star was found to vary with an amplitude as small as $0\fmm005$ (Fig.~\ref{fig10}).

	Other stars were found to exhibit periodic brightness decreases by $0\fmm015$. These light variations may be due to transits of ``hot Jupiter'' type 				exoplanets with the radii of  1/10 of the radius of the host stars and orbiting with less than one day-long periods~(\citet{Burdanov2013}).

	\subsection{Open Cluster NGC\,188}

\begin{figure}[tbp]
	\centering
	\includegraphics[width=0.9\columnwidth, clip]{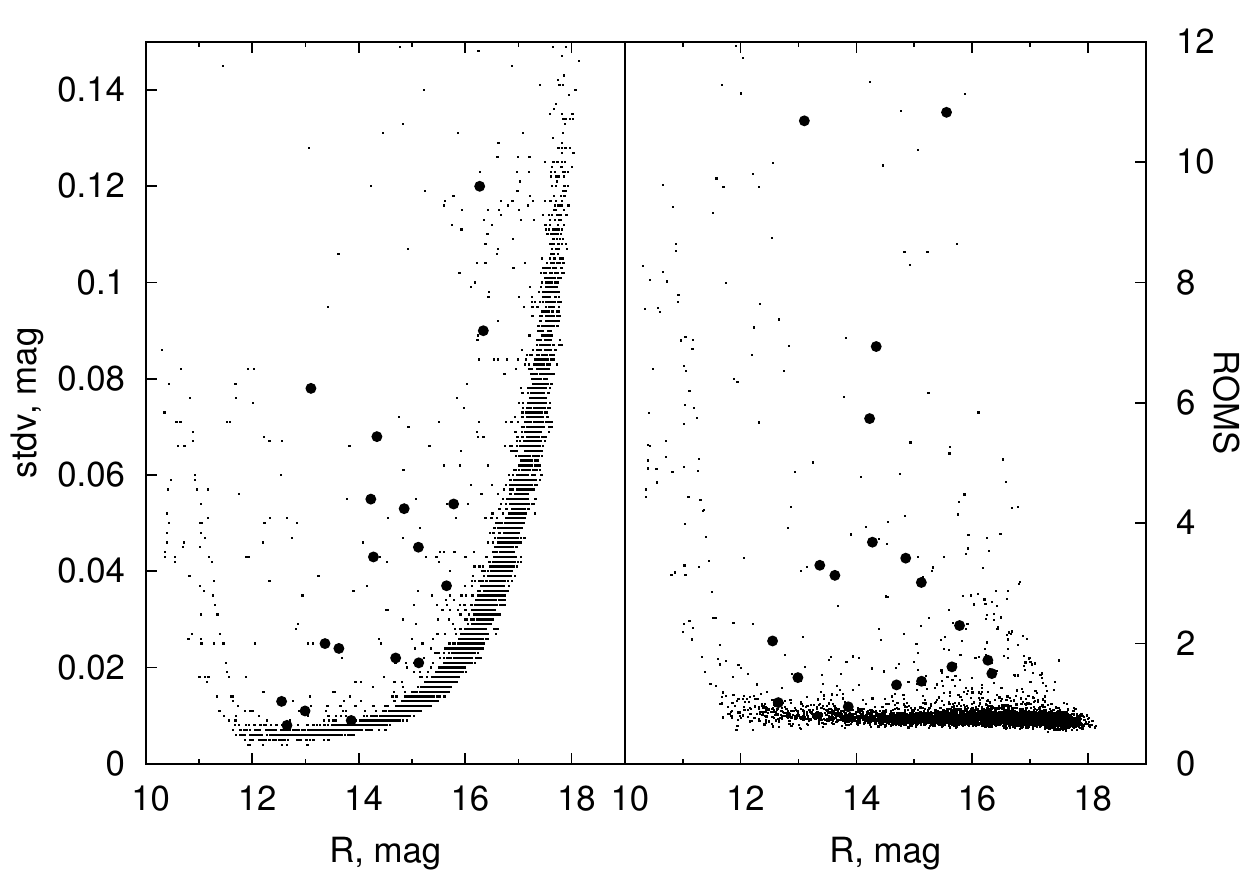}
	\caption{Standard deviations from the mean magnitude (left) and coefficients  ${\rm RoMS}$ (right) as a function of magnitude. The filled circles show 			variable stars.} 
	\label{fig11}
	\vspace{-10pt}
	\end{figure}

	The open cluster NGC\,188 was observed in the  $R$~and~$I$ filters with the \mbox{MASTER-II}-URAL telescope over five nights in March 2011. A total of  			400~frames were acquired, which were post-processed using  {\tt Astrokit} program. We performed aperture $R$-band photometry of 5513 stars with a precision
	ranging from $0\fmm006$ to $0\fmm05$ in a \mbox{$1\fdg5\times1\fdg5$} area. Although NGC\,188 is a well-studied cluster (more than  500~papers in the past 50 	years), our algorithm of the search for variable stars found 18~new variables~(\citet{Popov2013}). Figure~\ref{fig11} shows the standard deviation from
	the mean magnitude as a function of magnitude after reduction by our program (left) and the  ${\rm RoMS}$ coefficients for all stars (right). Variable stars 		are shown by the filled circles in both plots. As is evident from Fig.~\ref{fig11}, the standard deviation from the mean magnitude for variables often
	does not differ from the corresponding standard deviations for constant stars, whereas the ${\rm RoMS}$ coefficient allows more confident selection of 			candidate variable objects.

	\subsection{Observations of Exoplanet Transits}

	During the transits of even the largest  ``hot Jupiters''\ moving in close orbits across the disk of the host star, its brightness dips by about~$0\fmm01$. 		To confidently record the very fact of transit and determine its duration and mid-time, in the case of such brightness dips the resulting photometry should 		have the precision of about $0\fmm001$. Achieving the given precision is a difficult task for ground-based telescopes and requires special attention during 		differential photometry.

	Below we compare the results obtained with {\tt Astrokit} and the classical technique of differential photometry with a single comparison star, used to 			reduce the photometric observations of the transit of a ``hot Jupiter'' type exoplanet.

	Our $R$-band observations of the transit of the exoplanet \mbox{WASP-11\,b\,/\,HAT-P-10\,b} (\mbox{$R = 11\fmm01$}, \mbox{$J-H = 0.46$}, \mbox{$H-K = 0.14$}) 		were made with the \mbox{MASTER-II-Ural} telescope on December 10, 2012.

\begin{figure}[tbp!!!]
	\includegraphics[width=0.9\columnwidth, clip]{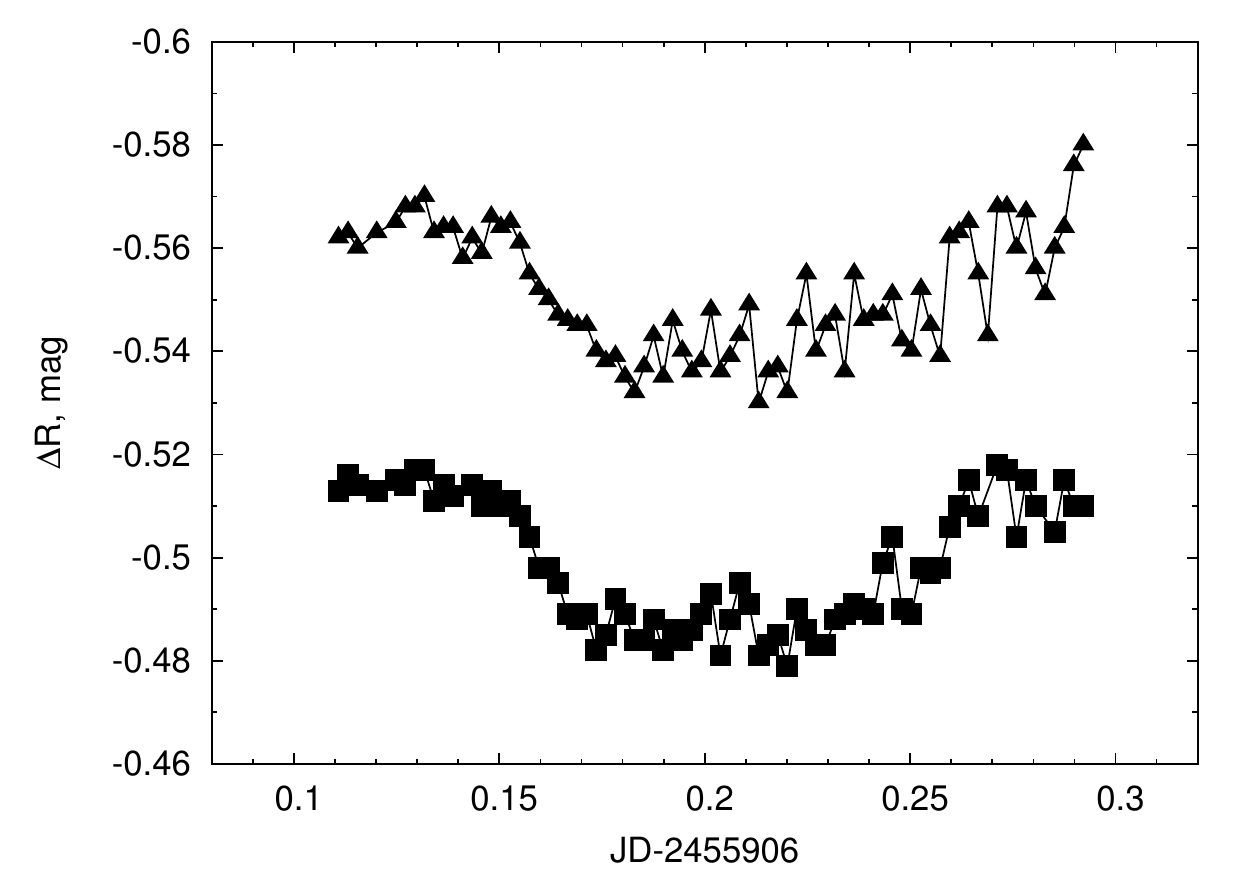}
	\caption{$R$-band light curve of the exoplanet
	\mbox{WASP-11\,b\,/\,HAT-P-10\,b} transit. The light curve obtained with a single comparison star is shown by the triangles, and that obtained using an 			ensemble of comparison stars---by the squares (it is shifted for better visualization).}
	\label{fig12}
	\end{figure}

	The light curve of the transit was obtained using the classical method of differential photometry, with a single comparison star 3UC-242-019494 (\mbox{$R = 12^{\rm 	m}$}, $J-H = 0.55$, \mbox{$H-K = 0.125$}), and a check star \mbox{3UC-242-019559} ($R = 11\fmm4$, $J-H = 0.175$,\linebreak \mbox{$H-K = 0.113$}). These stars 	are located within less than $10'$ from \mbox{WASP-11/HAT-P-10} ($R = 11\fmm7$) and have close magnitudes and color indices. The light curve obtained from 		magnitude difference between WASP-11/HAT-P-10 and the comparison star is shown by the triangles in Fig.~\ref{fig12}. The standard deviation of the 
	magnitude difference between the comparison and control star was $0\fmm006$.

	We used {\tt Astrokit} program with an ensemble of 11 stars to correct photometry for atmospheric transparency variation. The stars are located within  $6'$ 		and their magnitudes differ by no more than $2^{\rm m}$ from that of the star studied. We estimate the precision of corrected photometry ($0\fmm0039$) by the 	standard deviation from the mean magnitude for the control star 3UC-242-019559 mentioned above. The resulting precision is about a factor of 1.5 smaller that 	the standard deviation of the control-star magnitude in the case the classical method of differential photometry is used, thereby demonstrating the
	advantage of using ensembles of comparison stars. The resulting transit light curve is shown by the squares in Fig.~\ref{fig12}.

	This transit is remarkable by the fact that it was observed during total lunar eclipse. We clearly see that the scatter of data points increases toward the 		end of the transit. The standard deviation of the magnitude of the host star before the transit (the first 11~data points in the light curve) is equal to
	$0\fmm002$ and begins to increase with increasing sky background due to the egress of the Moon out of the Earth shadow. The standard deviation of host-star 		magnitude after the transit (the last 11 data points) is four times greater and equal to $0\fmm008$.

	\section{Conclusions}

	The use of a close ensemble of comparison stars while performing differential photometry allows the inhomogeneity of the data series, caused by local 				variations of atmospheric transparency and sky background, to be taken into account, and reduces the contribution of stellar scintillation  to the error 			budget of the resulting magnitudes~({\citet{Everett2001,Kornilov2012}}). If more than 10 reference stars in a close ensemble are used, the difference 				between their spectral types and that of the object studied becomes unimportant. However, to achieve the best precision, the magnitude difference should be 		small (it should not exceed $2^{\rm m}$) and ensemble stars should be chosen within $5'$--$7'$ of the program star.

	The resulting photometric precision after the application of  {\tt Astrokit} allows finding low-amplitude variables and study transits of ``hot Jupiters.\!''

	The use of robust median statistics is approved because of its higher stability against outliers. However, it does not guarantee against finding false 			variables. A considerable part of suspected variables prove to be constant stars within the photometric errors after further analysis. The number of 				candidate variables is usually equal to about  10\% of the total number of stars in the frame. The final criterion of the variability of a star is the
	visual inspection of its light curve.

	The source code of the program is available at \url{http://astro.ins.urfu.ru/en/node/1330}.

	\begin{acknowledgements}
	This work was supported by the Russian Foundation for Basic Research (projects No.~2-02-31095 and 14-02-31338). We are also grateful to Ekaterina Avvakumova 		for her assistance with the preparation of the paper and support. We also thank Kirill Ivanov for testing the program. This research has made use of the 			VizieR catalogue, SIMBAD database, and \mbox{Aladin} sky atlas, operated at CDS, Strasbourg, France.
	\end{acknowledgements}	

	\bibliographystyle{aa}
	\bibliography{astrokit.bib}
	
	\begin{flushright}
	{\it Translated by A.~Dambis}
	\end{flushright}

\end{document}